\documentclass[11pt,a4paper]{article}
\usepackage[utf8]{inputenc}
\usepackage[english]{babel}
\usepackage{amsmath,amssymb,graphicx}
\usepackage{booktabs}
\usepackage{geometry}
\usepackage{float} 
\usepackage{cite}  
\usepackage{url}

\title{\textbf{The JO\&Minna Framework: A Method to Interpret Depth Profiles of Activation}}
\author{Edil Mustafin}
\date{\today}

\begin{document}

\maketitle
\begin{abstract}
A novel analytical framework, designated as the JO\&Minna method, is presented for interpreting the depth profiles of residual activation induced by heavy-ion beams in thick structural materials. By combining the limiting fragmentation and factorization principles, the JO-factor formalism rigorously links the spatial distribution of projectile-like fragments to the range derivatives and electronic stopping powers. Furthermore, the Minna procedure establishes a mathematical link via profile differentiation and convolution analysis to reconstruct the original energy spectrum and energy spread of the fragments from the measured activity profiles. The methodology is benchmarked using experimental data for $^{46}\text{Sc}$ production from $^{238}\text{U}$ irradiation in aluminum, demonstrating a reliable and straightforward approach for extracting fragment production kinematics and energy distributions in accelerator materials research.
\end{abstract}
\section{Introduction}
The activation of accelerator components due to uncontrolled beam losses is a major factor limiting the intensity of high-energy hadron machines \cite{strasik2010}. The investigation of activation in various materials widely used in accelerators (such as copper, aluminum, and steel) irradiated by heavy-ion beams has been actively pursued at GSI Darmstadt over the past two decades \cite{strasik2010,katrik2022,katrik2015,katrik2019,mustafin2009_rad,mustafin2006_epac,fertman2007,strasik2008,strasik2010_ar,chetvertkova2011,kozlova2009,strasik2009_fluka,strasik2012,mustafin2003,mustafin2002}. 

The depth profile of the gamma-emitting radionuclide produced in thick targets by heavy ions exhibit characteristic features. It is common to distinguish between radionuclides produced from target nuclei (target-like fragments) and those produced from projectile nuclei (projectile-like fragments). Target-like fragments are typically distributed throughout the target thickness, whereas projectile-like fragments concentrate near and behind the primary ion range (i.e. behind the Bragg peak region). In this work, we focus on interpreting the activation profile of projectile-like fragments. Specifically, we examine the shape characteristics of the activity profile and propose a methodology to determine the energy spread of the fragment based on its profile.

\section{Characteristic Features of Projectile-Like Fragment Activation Profiles}

\begin{figure}[htbp]
    \centering
    \includegraphics[width=0.9\textwidth]{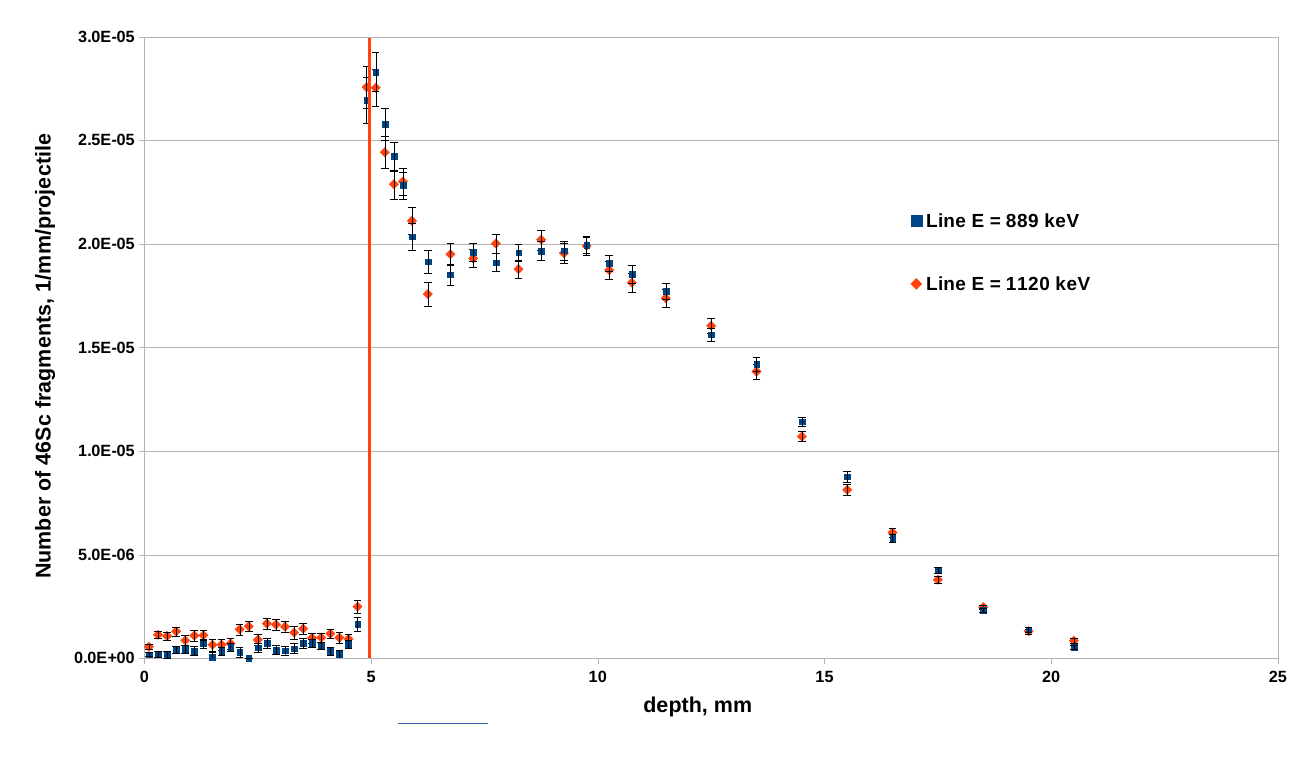}
    \caption{Activity profile of the $^{46}\text{Sc}$ fragment. The red vertical line indicates the range of the primary $^{238}\text{U}$ beam with an initial energy of 231.71 MeV/u. Markers of different colors correspond to the two main gamma-ray lines at 889 and 1120 keV.}
    \label{fig:profile_sc46}
\end{figure}

Let us examine the characteristic features of the activation profile induced in a thick target by projectile-like fragments, taking the nuclide $^{46}\text{Sc}$ produced by the irradiation of an aluminum target (99\% chemical purity) with a $^{238}\text{U}$ beam as a representative example, as shown in Figure~\ref{fig:profile_sc46}.

\begin{figure}[htbp]
    \centering
    \includegraphics[width=0.9\textwidth]{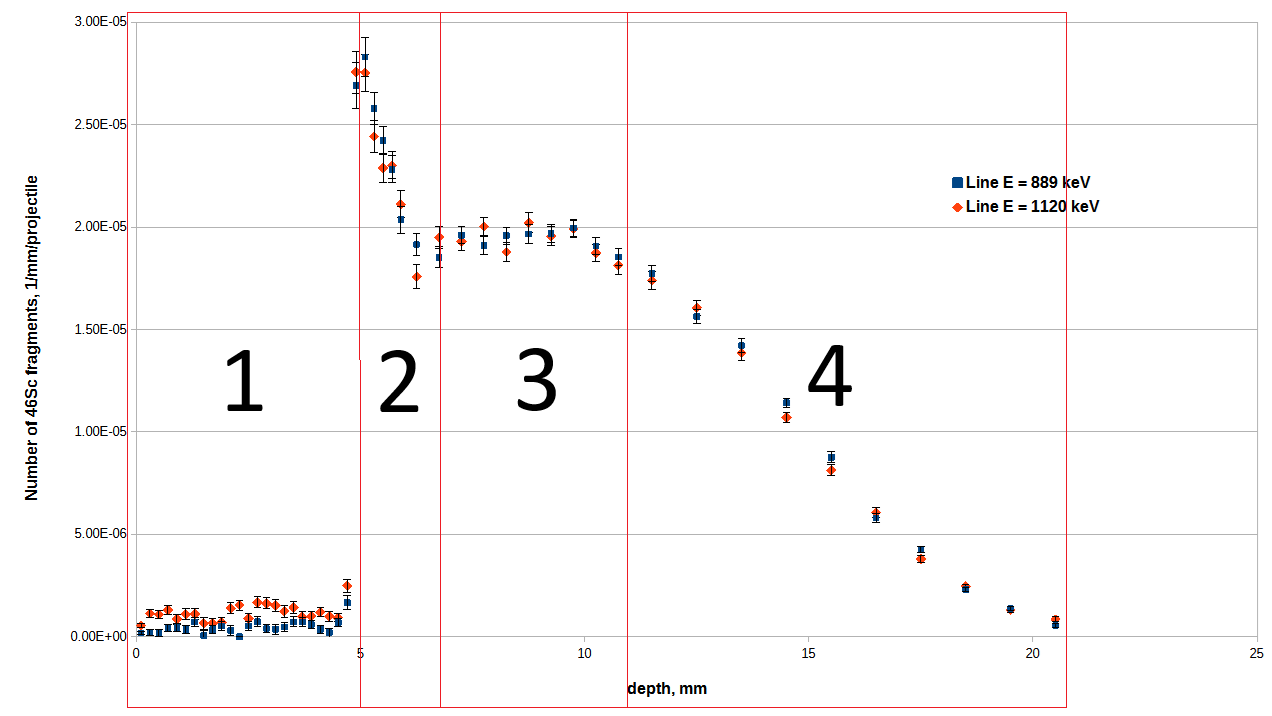}
    \caption{Characteristic regions (1 through 4) of the $^{46}\text{Sc}$ activity profile.}
    \label{fig:regions}
\end{figure}

Projectile-like fragments are characterized by the absence of induced activation up to the primary projectile range, designated as Region 1 in Figure~\ref{fig:regions}. Minor residual activity observed in this region is typically associated with heavy impurities (such as Fe and Cu) present in the aluminum target. Region 2 is a narrow zone adjacent to the projectile range, where the activity profile exhibits a prominent peak structure. Region 3 is characterized by a plateau-like behavior, while Region 4 displays a falling tail profile.

The choice of $^{46}\text{Sc}$ is motivated by two key factors: (1) this nuclide is sufficiently light to form an extended profile beyond the projectile range, making it particularly convenient for analysis, and (2) it has no radioactive isobars subject to $\beta$-decay or electron capture ($^{46}\text{Ca}$ and $^{46}\text{Ti}$ are stable nuclides), ensuring that its activity profile remains unaffected by the decay of precursor nuclides settled within the same target layer.

\section{Mechanism of Projectile-Like Fragment Activation Formation}
To understand the behavior of the projectile-like fragment activity profile, let us examine its formation mechanism in detail.

\begin{figure}[htbp]
    \centering
    \includegraphics[width=0.65\textwidth]{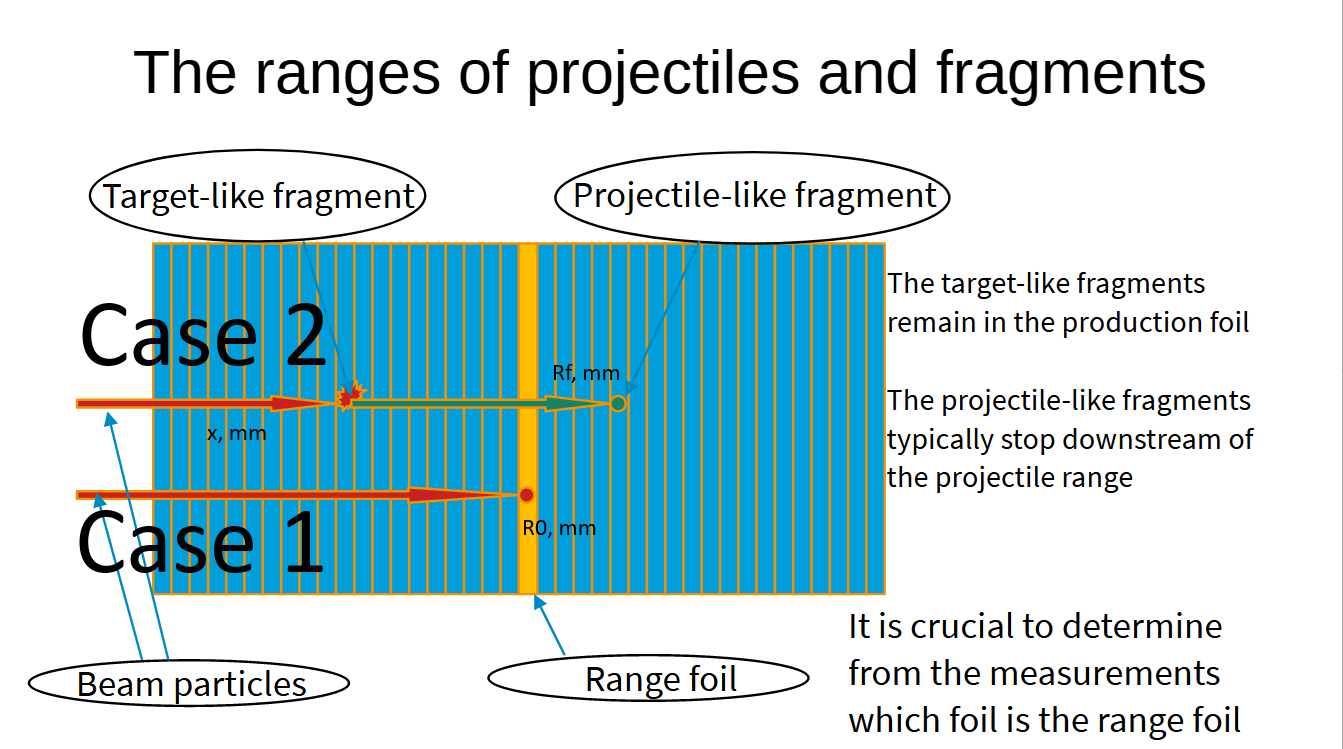}
    \caption{Mechanism of target-like and projectile-like fragment activation formation. Target-like fragments remain in the production foil, while projectile-like fragments typically stop downstream of the projectile range.}
    \label{fig:mechanism}
\end{figure}

A projectile entering the target may avoid nuclear interactions with target nuclei, undergo ionization energy loss, and come to rest at a depth corresponding to its range. This scenario is illustrated as \textbf{Case 1} in Figure~\ref{fig:mechanism}. The foil in which the unreacted projectile stops is referred to as the \textit{range foil} (highlighted in Figure~\ref{fig:mechanism})[cite: 2].

\textbf{Case 2} describes a situation where the projectile undergoes a nuclear reaction with a target nucleus at a certain depth after losing part of its initial energy via electronic stopping. The foil where the reaction occurs is referred to hereafter as the \textit{reaction foil}. In this case, the resulting projectile-like fragment retains essentially the same velocity per nucleon as the projectile just before the collision. Since the fragment is typically lighter and has a lower nuclear charge than the primary projectile, it experiences a lower electronic stopping power. Consequently, the fragment generally stops deeper in the target than the primary projectile range, which accounts for the absence of activation in Region 1 of Figure~\ref{fig:regions}.

To demonstrate how to interpret the activity profile behavior in Regions 2 and 3, let us examine Case 2 more closely. Let $x$ denote the distance from the front surface of the target (entry point) to the reaction point, and let $E_p$ be the kinetic energy of the projectile at that point. Obviously, $E_p$ equals the initial energy minus the energy lost through electronic stopping. Furthermore, let $R_f$ denote the range of the fragment with energy $E_p$ (assuming the fragment emerges with the same energy per nucleon as the projectile entering the interaction) from the interaction point $x$ to its complete stop. Let $R_0$ denote the range corresponding to the initial projectile energy. Finally, $r_f$ represents the \textit{residual range of the fragment}: the distance from $R_0$ to the fragment stopping point.

The following geometrical relations hold among these quantities:
\begin{align}
    x(E_p) &= R_0 - R_p(E_p) \label{eq:x_ep} \\
    r_f(E_p) &= x(E_p) + R_f(E_p) - R_0 = R_f(E_p) - R_p(E_p) \label{eq:rf_ep}
\end{align}
Here, we emphasize that: (a) the residual fragment range $r_f$ is measured from the primary projectile range $R_0$ (rather than from the target surface), and (b) the projectile range $R_p$ and fragment range $R_f$ are evaluated for the identical energy $E_p$ at which the nuclear reaction took place.

The values of $R_p$ and $R_f$ for various energies $E_p$, calculated using the \textbf{ATIMA} code \cite{atima}, along with the respective electronic stopping powers, are summarized in Table~\ref{tab:jofactor_table}.

\begin{table}[htbp]
\centering
\caption{Energy loss, ranges for $^{46}\text{Sc}$ fragment and $^{238}\text{U}$ projectile, and corresponding JO-factor values.}
\label{tab:jofactor_table}
\small
\begin{tabular}{ccccccc}
\toprule
$E_p$ & \multicolumn{2}{c}{$^{46}\text{Sc}$ Fragment} & \multicolumn{2}{c}{$^{238}\text{U}$ Projectile} & & \\
\cmidrule(r){2-3} \cmidrule(l){4-5}
(MeV/u) & $dE/dx$ ($\frac{\text{MeV}}{\text{g/cm}^2}$) & $R_f$ (mm) & $dE/dx$ ($\frac{\text{MeV}}{\text{g/cm}^2}$) & $R_p$ (mm) & $r_f$ (mm) & JO-factor \\
\midrule
231.71 & $1.44 \times 10^3$ & 16.17 & $2.68 \times 10^4$ & 4.98 & 11.19 & 0.385 \\
200    & $1.57 \times 10^3$ & 12.64 & $2.88 \times 10^4$ & 4.00 &  8.65 & 0.393 \\
170    & $1.74 \times 10^3$ &  9.62 & $3.13 \times 10^4$ & 3.13 &  6.49 & 0.404 \\
150    & $1.89 \times 10^3$ &  7.78 & $3.34 \times 10^4$ & 2.60 &  5.18 & 0.414 \\
120    & $2.20 \times 10^3$ &  5.32 & $3.77 \times 10^4$ & 1.87 &  3.45 & 0.432 \\
100    & $2.50 \times 10^3$ &  3.89 & $4.17 \times 10^4$ & 1.43 &  2.46 & 0.450 \\
80     & $2.94 \times 10^3$ &  2.66 & $4.71 \times 10^4$ & 1.04 &  1.62 & 0.477 \\
50     & $4.15 \times 10^3$ &  1.21 & $5.96 \times 10^4$ & 0.55 &  0.66 & 0.563 \\
20     & $7.87 \times 10^3$ &  0.28 & $8.92 \times 10^4$ & 0.19 &  0.09 & 0.840 \\
10     & $1.21 \times 10^4$ &  0.10 & $1.01 \times 10^5$ & 0.10 &  0.00 & 1.631 \\
\bottomrule
\end{tabular}
\end{table}

\section{The JO-Factor Formalism}
To interpret the activation depth profiles, we introduce the JO $\&$ Minna framework. Here, JO refers to the Junction-Offset formalism designed to properly account for the characteristic peak-and-plateau structures, while Minna stands for Monotonic Interpolation $\&$ Numerical Normalization Algorithm, which handles the stable reconstruction of the underlying fragment energy distributions without amplifying experimental noise.

To determine how the fragments are distributed across the foils, we apply two fundamental principles:
\begin{enumerate}
    \item \textbf{Limiting Fragmentation Principle:} The total nucleus-nucleus reaction cross-section is virtually independent of collision energy (at energies above tens of MeV/u). This principle underlies the Kox parameterization \cite{kox1987} valid for $E_p$ from 30 to 300 MeV/u.
    \item \textbf{Factorisation Principle:} The production cross-section of a specific fragment is also independent of energy, as utilized in the EPAX parameterization \cite{summerer2000}.
\end{enumerate}

From these principles, if Region 1 consists of foils of equal thickness, the number of nuclear reactions and produced fragments per foil remains constant (except near the range foil where the beam slows below tens of MeV/u). 

Let $dN$ be the number of a specific projectile-like fragment produced in a reaction foil of thickness $dR_p$, then all these fragments would settle in the target in a layer of thickness $dr_f \neq dR_p$. Here, $dR_p$ represents the difference in projectile ranges corresponding to the projectile energy at the entrance ($E_p$ ) and the exit ($E_p+dE_P$) of the reaction foil, while $dr_f$ is the difference in fragment residual ranges corresponding to the same energy values. Obviously, $dE_p/dR_p = (dE/dx)_p$ and $dE_p/dR_f = (dE/dx)_f$, where the indices $p$ and $f$ at $dE/dx$ refer to projectile and fragment. 

The quantity we usually derive from the measurement of the gamma-spectra is $dN/dr_f$:
\begin{equation}
    \frac{dN}{dr_f} = \frac{dN}{dR_f - dR_p} = \frac{dN/dR_p}{\frac{dR_f}{dR_p} - 1} \label{eq:dN_drf}
\end{equation}
$dN/dR_p$ is the number of the projectile fragment produced in the reaction foil and it is constant across Region 1. Thus, the measured quantity $dN/dr_f$ depends on the derivative of ranges only:
\begin{equation}
    \frac{dR_f}{dR_p} = \frac{A_f}{A_p}\frac{(dE/dx)_p}{(dE/dx)_f} \label{eq:drf_drp}
\end{equation}
where $(dE/dx)_p$ and $(dE/dx)_f$ are the stopping powers of the projectile and fragment in the reaction foil, $A_p$ and $A_f$ are the respective mass numbers. 

Finally, $dN/dr_f$ is proportional to the \textbf{JO-factor}:
\begin{equation}
    \text{JO-factor} = \left( \frac{A_f}{A_p}\frac{(dE/dx)_p}{(dE/dx)_f} - 1 \right)^{-1} \label{eq:jofactor}
\end{equation}

As $E_p$ and consequently the fragment residual range increase, the JO-factor approaches an asymptotic limit determined by the charge-mass ratio:
\begin{equation}
    \frac{1}{\frac{A_{^{46}{Sc}}}{A_{^{238}U}}\left( \frac{Z_{^{238}U}}{Z_{^{46}{Sc}}} \right)^2 - 1} = \frac{1}{\frac{46}{238}\left( \frac{92}{21} \right)^2 - 1} \approx 0.369 \label{eq:asymptotic}
\end{equation}

\begin{figure}[htbp]
    \centering
    \includegraphics[width=0.8\textwidth]{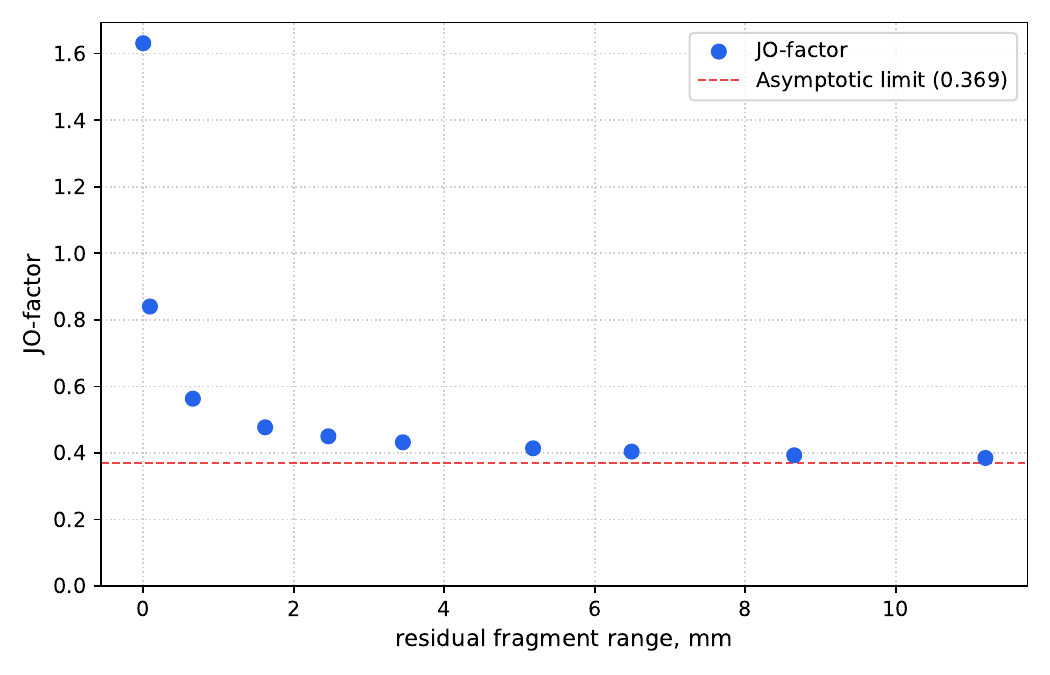}
    \caption{Dependence of the JO-factor on the residual range of the projectile-like fragment. The horizontal red line represents the asymptotic limit of 0.369.}
    \label{fig:jofactor_plot}
\end{figure}

Table~\ref{tab:jofactor_table} lists the JO-factor values for various projectile and fragment energies (assuming that the fragment energy after the reaction equals the projectile energy before the reaction), while Figure~\ref{fig:jofactor_plot} shows the dependence of the JO-factor on the residual range of the fragment (i.e., the horizontal axis represents the values from the penultimate column of Table~\ref{tab:jofactor_table}, and the vertical axis represents the values from the last column).

Note that zero on the horizontal axis of Figure~\ref{fig:jofactor_plot} corresponds to the primary projectile range. The high JO-factor values immediately behind the range foil and the gradual decrease of the JO-factor toward the constant $0.369$ explain the behavior of the projectile-like fragment activity profile in Regions 2 and 3 in Figure~\ref{fig:regions}. One could say that Region 2 in Figure~\ref{fig:regions} corresponds to a JO-factor from 0 to about 2 mm ($r_f=2$ mm corresponds to $E_p \approx 90$ MeV/u) in Figure~\ref{fig:jofactor_plot}, and the plateau of Region 3 in Figure~\ref{fig:regions} corresponds to a JO-factor from 2 mm onward in Figure~\ref{fig:jofactor_plot}.

\section{Region 4: Energy Spectrum of Fragments and the Formation of the Falling Tail}
\label{sec:case4}

To explain the activity drop in Region 4 (the falling tail of the profile), it is necessary to account for the actual energy spectrum of the produced fragments. The analysis of this mechanism can be naturally divided into several successive steps, moving from an idealized situation to the realistic distribution within the target.

\begin{enumerate}
    \item \textbf{Monogenetic Fragment Production and the Ideal Step-like Function ($S_{\text{ideal}}$):}
    Let us consider a group of foils at the beginning of the target sufficiently far away from the range foil to assume that the electronic stopping power is almost constant over the foils (or in other words JO-factor is almost constant).
    In accordance with the principles of limiting fragmentation and factorization, the nuclear interaction cross-section is independent of energy at high energies (above tens of MeV/u), meaning that the projectile fragments are produced \textbf{uniformly across all the foils under consideration}. If we assume the fragments are produced with no energy spread they rest uniformly distributed over some region beyond the projectile range. Taking into account the stopping kinetics and the JO-factor, it ideally forms a \textbf{step-like distribution $S_{\text{ideal}}(x)$}, as shown in Figure~\ref{fig:ideal_Sx}.
    
    \begin{figure}[htbp]
        \centering
        \includegraphics[width=0.8\textwidth]{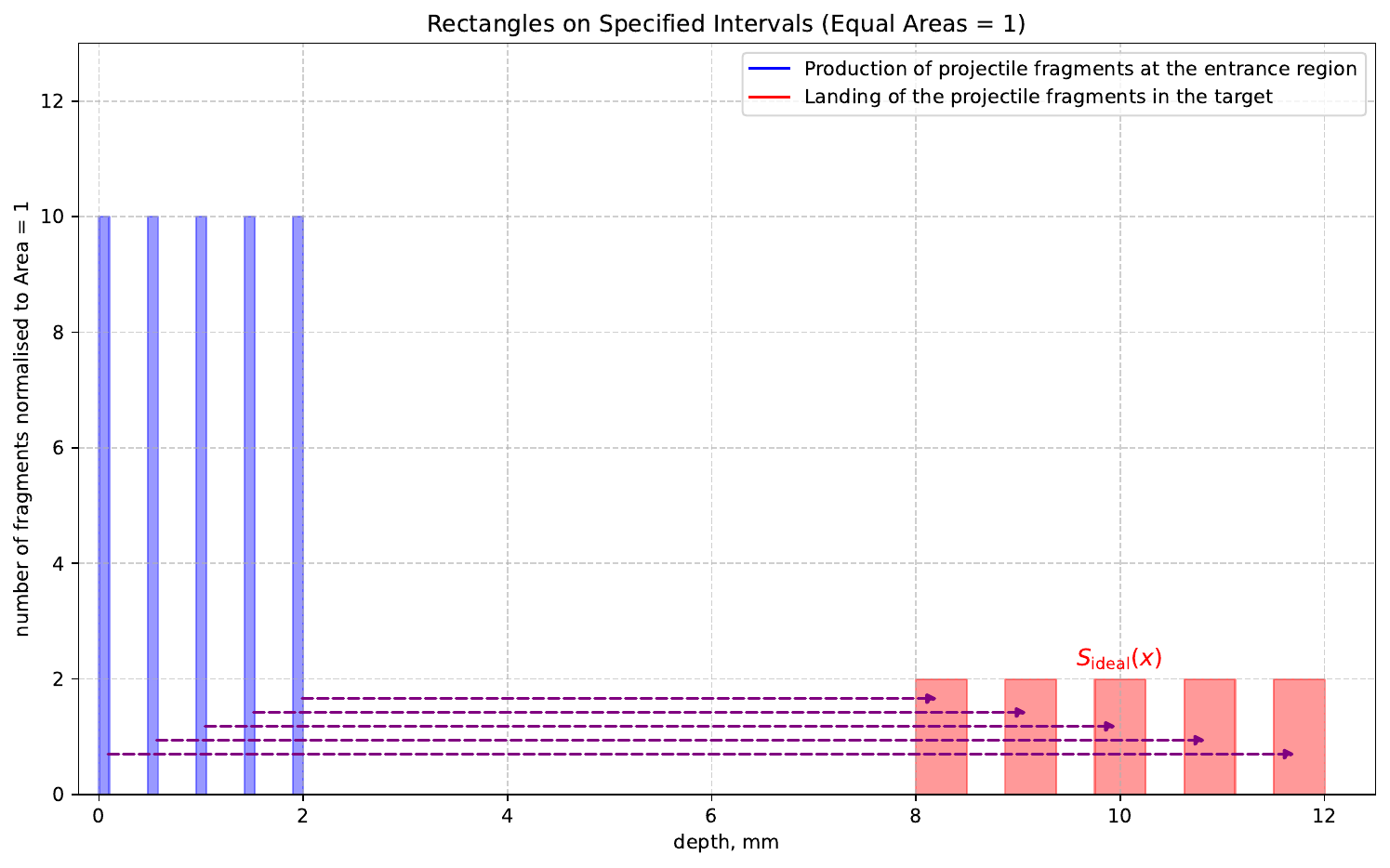}
        \caption{Family of fragment rectangular distributions: blue rectangles represent uniform production of projectile fragments in different reaction foils, red rectangles represent distribution of projectiles stopped in the target assuming no energy spread of fragments at production.}
        \label{fig:ideal_Sx}
    \end{figure}

    \item \textbf{Energy Spread in Production Foils:}
    Now let's take into account that the fragments in each production foil are not produced mono-energetically, but with an energy spread around the reaction energy. Born with such an energy spread, these fragments begin to slow down in the material. Since the fragment range $R_f(E)$ depends almost linearly on energy at high energies, the fragment energy spread in each foil is directly and proportionally mapped into a \textbf{corresponding depth distribution in the target} beyond the projectile range. For illustrative purposes let's assume Gaussian energy spread of fragments at production. If we trace this process discretely, for each target layer we obtain a family of shifted Gaussian curves, as illustrated in Figure~\ref{fig:gaussians}.
    
    \begin{figure}[htbp]
        \centering
        \includegraphics[width=0.8\textwidth]{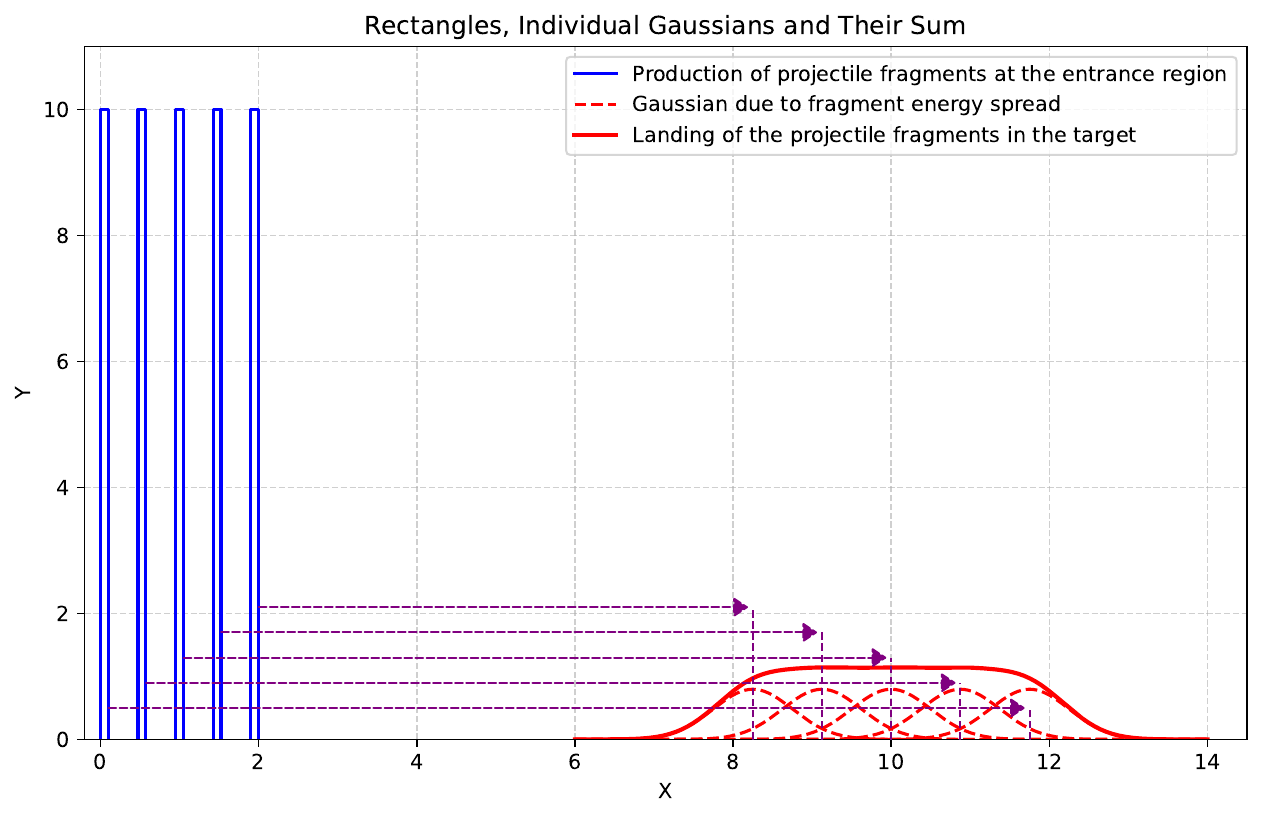}
        \caption{Family of fragment Gaussian distributions formed due to the energy spread at production.}
        \label{fig:gaussians}
    \end{figure}

    \item \textbf{Superposition of Contributions and Convolution with a Gaussian Kernel:}
    Because nuclear reactions occur continuously throughout the entire thickness of the target section under consideration, each elementary layer adds its own Gaussian contribution to the fragment residual range profile. The superposition (integration) of these distributions is a convolution of the $S_{\text{ideal}}(x)$ with a Gaussian Kernel. The ideal step-like distribution $S_{\text{ideal}}(x)$, reflecting the uniform distribution of sources at the beginning of the target, undergoes physical smoothing due to the energy spread of the fragments (which act as a Gaussian kernel with an effective distribution width $\sigma$) Figure~\ref{fig:summed_gaussian_curve}:

    \begin{equation}
        S_{\text{real}}(x) = \int_{-\infty}^{\infty} S_{\text{ideal}}(x') \, \frac{1}{\sigma\sqrt{2\pi}} \exp\left(-\frac{(x - x')^2}{2\sigma^2}\right) dx' \label{eq:convolution}
    \end{equation}

    \begin{figure}[htbp]
        \centering
        \includegraphics[width=0.8\textwidth]{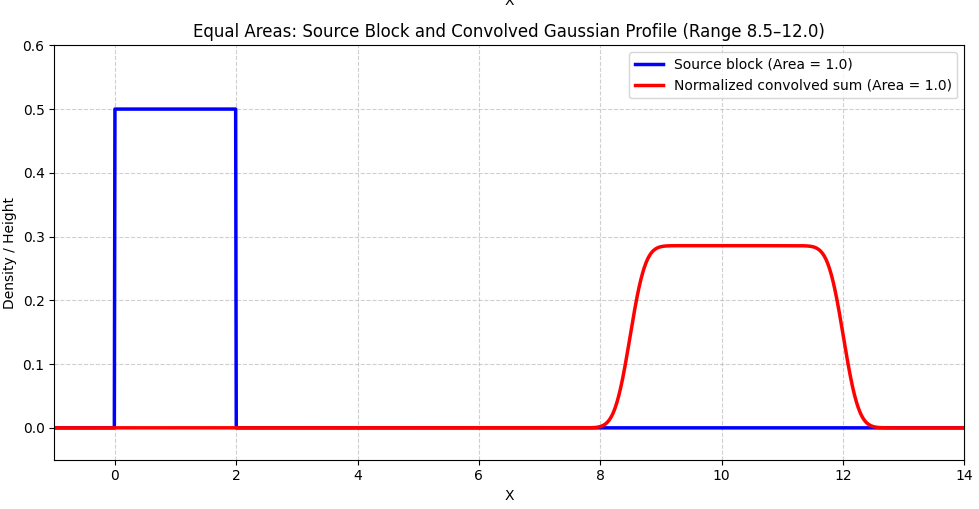}
        \caption{All rectangles representing the uniform source distribution and the resulting total summed curve formed by the Gaussian profiles.}
        \label{fig:summed_gaussian_curve}
    \end{figure}

    It is precisely this convolution operation that smooths the right-hand boundary of the ideal step-like profile, forming the characteristic falling tail in Region 4. Furthermore, taking the derivative of the resulting profile function $S(x)$ with respect to the coordinate yields a characteristic dipole structure (consisting of positive and negative peaks at the boundaries, as shown in Figure~\ref{fig:dipole_plot}).

    \begin{figure}[htbp]
        \centering
        \includegraphics[width=0.85\textwidth]{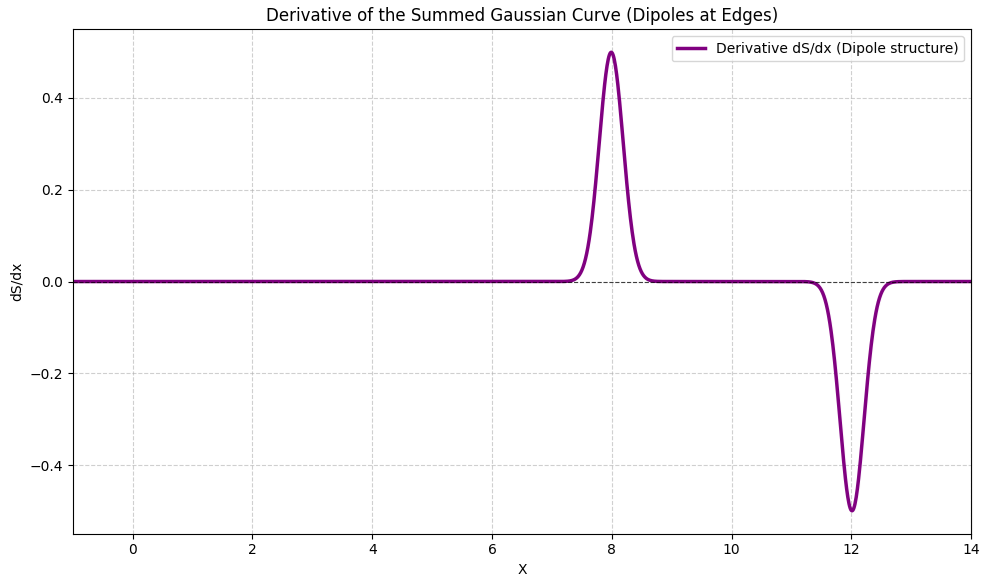}
        \caption{Derivative of the summed Gaussian curve ($dS/dx$), demonstrating the dipole structure at the edges.}
        \label{fig:dipole_plot}
    \end{figure}

    This derivative directly reveals the underlying fragment energy spread. Note that the $\sigma$ found in this way has a dimension of space (millimeters). In order to convert it into the width of energy distribution one has to multiply it to $dE/dx$ at the production foil. According to our assumption $dE/dx$ is the same in all foils under consideration (the assumption is valid for the regions where JO-factor is about constant): $\sigma_E = \sigma \times (dE/dx)_f$.
    The proposed approach allows linking the measured shape of this tail with the actual energy distribution of the fragments.
\end{enumerate}

\section{Rigorous Mathematical Approach}
\label{sec:RMA}
\begin{figure}[htbp]
    \centering
    \includegraphics[width=0.85\textwidth]{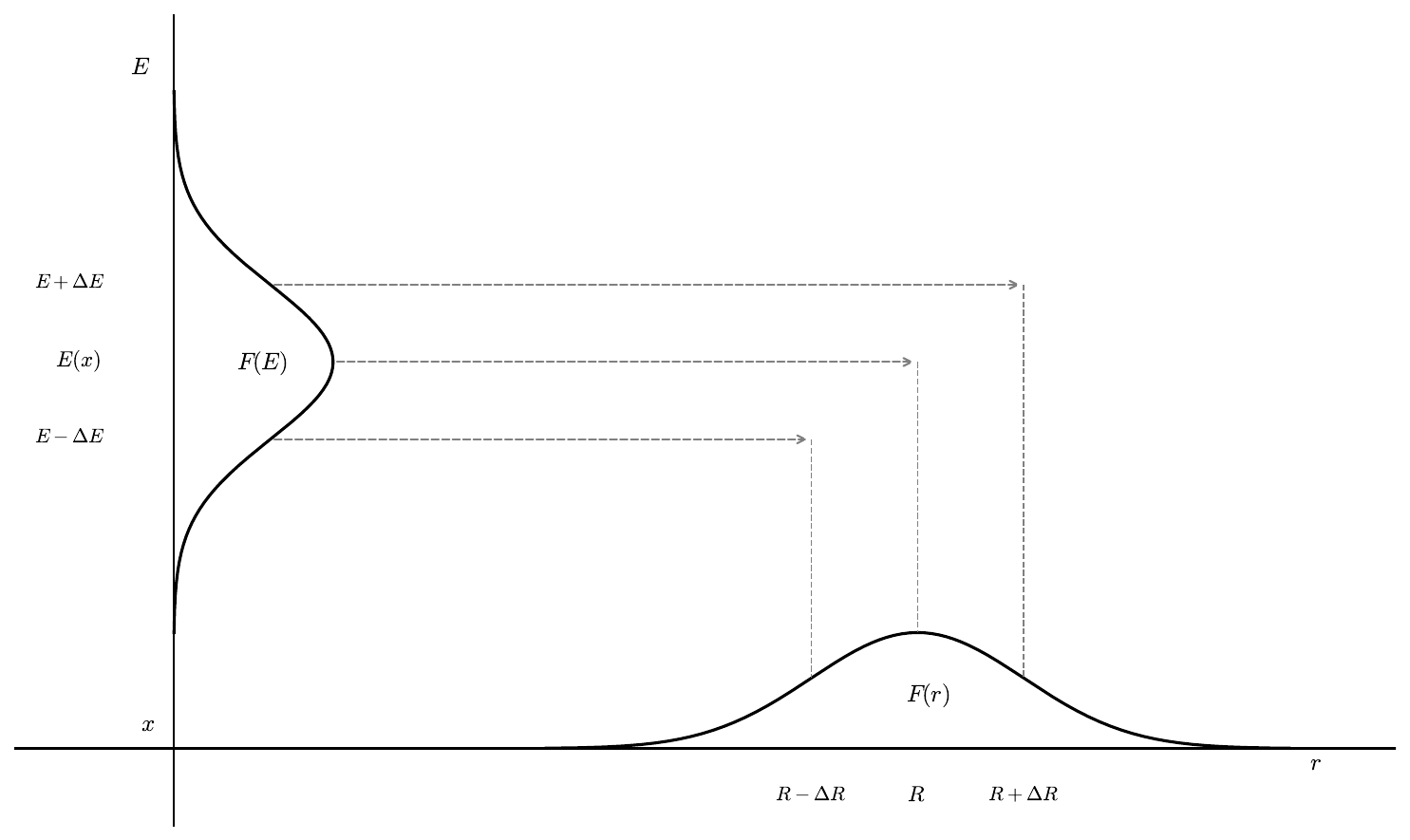}
    \caption{Schematic representation of the relationship between the fragment energy distribution $F(E)$ in a layer at coordinate $x$ and their spatial distribution $F(r)$.}
    \label{fig:E_R_scheme}
\end{figure}

Let the projectile fragments be produced in a layer at coordinate $x$ with an energy spread described by the function $F(E)$ (for clarity, a Gaussian dependence is chosen as an example in Figure~\ref{fig:E_R_scheme}). 

We assume the distribution to be sufficiently narrow so that the dependence of $\mathrm{d}E/\mathrm{d}x$ on energy in the vicinity of $E \pm \Delta E$ can be neglected. In this approximation, the form of the functional dependence is preserved: if the energy distribution is Gaussian with a width $\sigma_E$, the spatial distribution of fragments will also have a Gaussian form. Fragments with energy $E$ stop at a depth $r = x + R_f(E)$, where $R_f(E)$ is the range of the fragment with energy $E$. Accordingly, the spectrum components with energies $E \pm \Delta E$ are localized at depths $R \pm \Delta R$, with the spatial width related to the energy distribution width by the relation $\sigma_r = \sigma_E / (\mathrm{d}E/\mathrm{d}x)_f$ (or in increments: $\Delta R = \Delta E / (\mathrm{d}E/\mathrm{d}x)_f$), and the range-energy dependence $R(E)$ in the considered range can be considered linear:
\begin{equation}
    R_f(E) = R_{f,0} \frac{E}{E_0},
    \label{eq:linear_range}
\end{equation}
where $R_{f,0}$ is the range of a fragment with energy $E_0$ born at the target surface.

This approximation is equivalent to the constancy of the stopping power over the corresponding energy interval. We are interested in a narrow group of fragments generated at the very beginning of the target, which stop in the target near a characteristic depth differing from the maximum fragment range $R_{f,0}$ by a few distribution variances ($\pm \sigma_r$).

The measured physical quantity in the experiment is the spatial distribution of the fragment concentration $\mathrm{d}N/\mathrm{d}r$ in the layer at coordinate $r$. To find it, one must integrate the contributions from all fragments born in various target layers $x$ (along the entire path of the primary projectile up to its total range $R_{0,\text{p}}$) that settle at a given depth $r$ after slowing down. 

Let fragments distributed in energy according to the function $F(E)$ be produced in an elementary target layer $\mathrm{d}x$ with a generation rate $Y$. We assume that the energy distribution function doesn't depend on the production coordinate $x$ over the range under consideration.

Each such fragment, possessing energy $E$, travels a path equal to its range $R_f(E)$ and stops at depth $r = x + R_f(E)$. The strict condition that a fragment born at point $x$ with energy $E$ contributes to the concentration at a fixed depth $r$ is given by means of the Dirac delta function:
\begin{equation}
    \frac{\mathrm{d}N}{\mathrm{d}r}(r) = Y \int_{0}^{R_{0,\text{p}}} \mathrm{d}x \int_{0}^{\infty} \mathrm{d}E \, F(E) \cdot \delta\left(r - x - R_f(E)\right),
    \label{eq:delta_integral}
\end{equation}

Within the framework of our approximations, the range-energy dependence is considered linear (see \eqref{eq:linear_range}): $R_f(E) = R_{f,0} {E}/{E_0}$ (which is equivalent to the constancy of the stopping power $(\mathrm{d}E/\mathrm{d}r)_f = \text{const}$). Using the property of the delta function under variable substitution and switching to integration over energy taking into account the scale factor (the transformation Jacobian $\left| \mathrm{d}E/\mathrm{d}r \right|_f$), we evaluate the inner integral in expression \eqref{eq:delta_integral} over energy. As a result, we obtain the final expression in the form of a convolution of the initial generation spectrum with the spatial profile:
\begin{equation}
    \frac{\mathrm{d}N}{\mathrm{d}r}(r) = Y \cdot \left| \frac{\mathrm{d}E}{\mathrm{d}r} \right|_f \int_{0}^{R_{0,\text{p}}} F\left(\frac{E_0}{R_{f,0}}(r - x)\right) \mathrm{d}x.
    \label{eq:convolution_result}
\end{equation}

To reconstruct the initial energy spectrum from the experimentally measured spatial distribution, we apply the differentiation operator with respect to coordinate $r$ to both parts of equation \eqref{eq:convolution_result}. Replacing the derivative with respect to the parameter $r$ by the derivative with respect to the production coordinate $x$ via the relation $\frac{\partial}{\partial r} = -\frac{\partial}{\partial x}$, we rewrite the integral in the form of a derivative:
\begin{equation}
    \frac{\mathrm{d}}{\mathrm{d}r} \left( \frac{\mathrm{d}N}{\mathrm{d}r}(r) \right) = - Y \cdot \left| \frac{\mathrm{d}E}{\mathrm{d}r} \right|_f \int_{0}^{R_{0,\text{p}}} \frac{\partial}{\partial x} \left[ F\left(\frac{E_0}{R_{f,0}}(r - x)\right) \right] \mathrm{d}x.
    \label{eq:derivative_step}
\end{equation}

Applying the Newton-Leibniz formula to expression \eqref{eq:derivative_step}, we proceed to the function values at the boundaries of the integration interval:
\begin{equation}
    \frac{\mathrm{d}}{\mathrm{d}r} \left( \frac{\mathrm{d}N}{\mathrm{d}r}(r) \right) = - Y \cdot \left| \frac{\mathrm{d}E}{\mathrm{d}r} \right|_f \left[ F\left(\frac{E_0}{R_{f,0}}(r - x)\right) \right]_{x=0}^{x=R_{0,\text{p}}}.
    \label{eq:neiton_leibniz}
\end{equation}

Substituting the limits, we obtain the difference between the function values at the upper and lower thresholds:
\begin{equation}
    \frac{\mathrm{d}}{\mathrm{d}r} \left( \frac{\mathrm{d}N}{\mathrm{d}r}(r) \right) = Y \cdot \left| \frac{\mathrm{d}E}{\mathrm{d}r} \right|_f \left[ F\left(\frac{E_0}{R_{f,0}} r\right) - F\left(\frac{E_0}{R_{f,0}}(r - R_{0,\text{p}})\right) \right].
    \label{eq:limits_expansion}
\end{equation}

We are interested in the region of the activation profile tail, i.e., the region of $r$ close to $R_{f,0}$; in this region, only fragments generated near the target surface $x=0$ contribute to the profile, while the contribution of the upper limit is physically zero:
\begin{equation}
    \frac{\mathrm{d}}{\mathrm{d}r} \left( \frac{\mathrm{d}N}{\mathrm{d}r}(r) \right) \propto F\left(\frac{E_0}{R_{f,0}} r\right).
    \label{eq:derivative_result}
\end{equation}
The obtained expression \eqref{eq:derivative_result} demonstrates a direct proportionality between the derivative of the measured concentration profile and the initial energy spectrum of the fragments $F(E)$.

\section{Minna-Procedure: A Method for Extracting the Fragment Energy Distribution from the Activity Profile}
\label{sec:Minna}

As it was shown in the previous section, the derivative of the measured activity profile function in Region 4 with respect to the depth coordinate $x$ yields the original energy distribution function of the fragments, up to a normalization scale factor. 

Although a Gaussian shape was assumed for the energy spectrum above as an illustrative example, the actual energy distribution in practice can take any arbitrary form. Crucially, the treatment within the JO-factor formalism remains entirely unaffected: only the kernel function (the Gaussian) within the convolution integral is replaced by the actual energy distribution function. 

In order to get a smooth derivative, let us fit the measured points with the function $S(x)$ from Equation~\ref{eq:convolution}.
    \begin{figure}[htbp]
        \centering
        \includegraphics[width=0.85\textwidth]{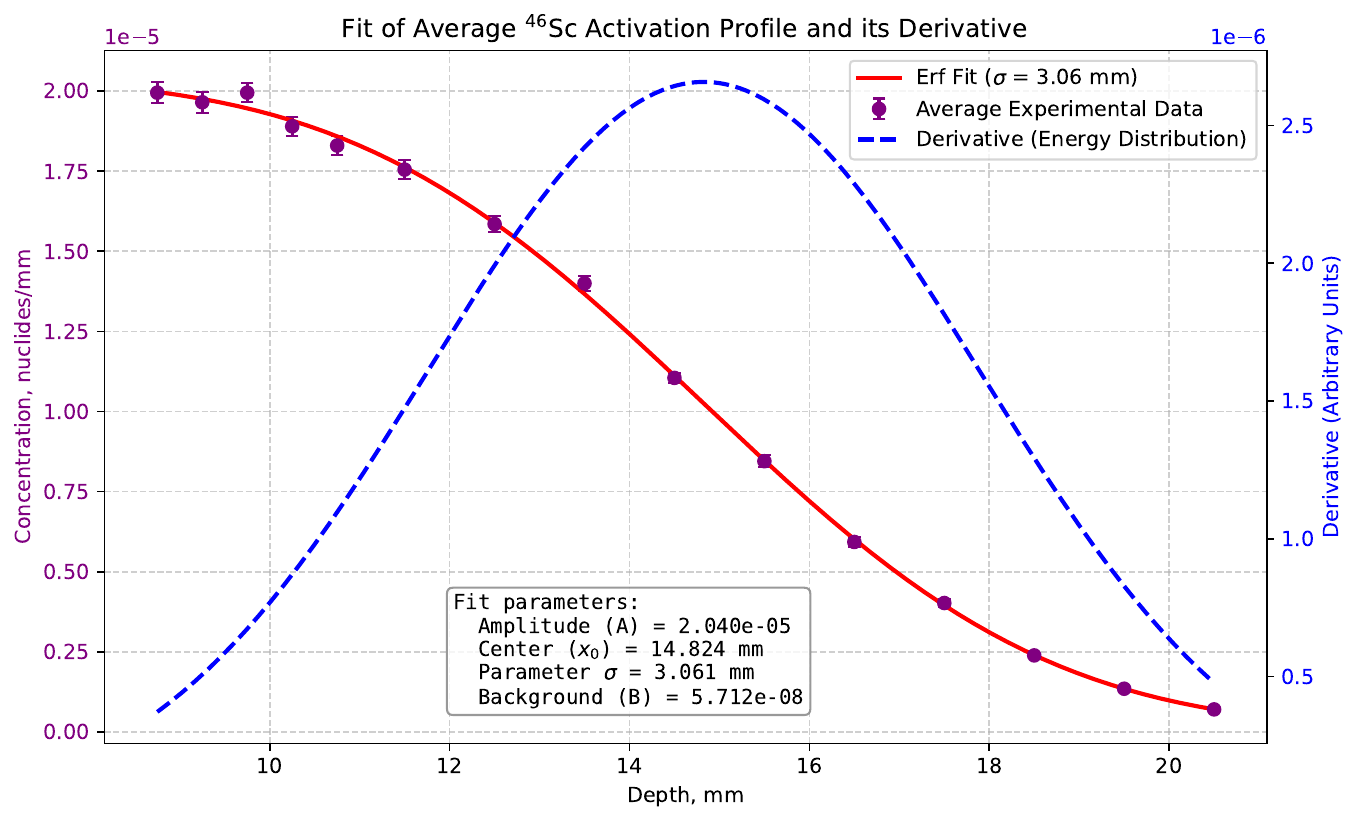}
        \caption{Fit of the experimental curve.}
        \label{fig:46Sc_Erf_fit}
    \end{figure}
    As it is seen in Figure~\ref{fig:46Sc_Erf_fit} the width is $\sigma=3.06$ mm. Or, converting to energy distribution:
    
     \begin{equation}
        \sigma_{\text{E}} = \sigma \times (dE/dx)_f = 3.06 \times 1.44\times 10^3 \times 2.76 \times 0.1 = 415 \text{MeV}
        \label{eq:sigma_E}
    \end{equation}
    Here $\sigma = 3.06$ mm, $(dE/dx)_f=1.44\times 10^3$ MeV/(g/cm$^2$), $\rho_\text{Al}=2.76$ g/cm$^3$ is the density of the target material, 0.1 converts cm to mm. More convenient is the specific unit MeV/u: $\sigma = 415/46 \approx $ 9 MeV/u or $\text{FWHM} = 2\sqrt{2\ln 2} \cdot \sigma \approx 2.35482 \cdot \sigma = 2.35482 \cdot 9 \approx 21$ MeV/u.
    
    The center of the derivative corresponds to the range of $^{46}Sc$ $R_f=$ 14.8 mm. ATIMA calculation shows that the energy $E_c$ corresponding to such range of $^{46}Sc$ in aluminum with density 2.76 g/cm$^2$ is $E_c=220$ MeV/u.
    
    Concluding, the Minna-Procedure estimates from the measurements that $^{238}U$ with the energy $E_p=231.71$ MeV/u interacting with $^{27}Al$ nuclei produces projectile fragments $^{46}Sc$ with the central energy 220 MeV/u and the energy spread around it with FWHM $\approx$ 21 MeV/u. 

\begin{figure}[htbp]
        \centering
        \includegraphics[width=0.85\textwidth]{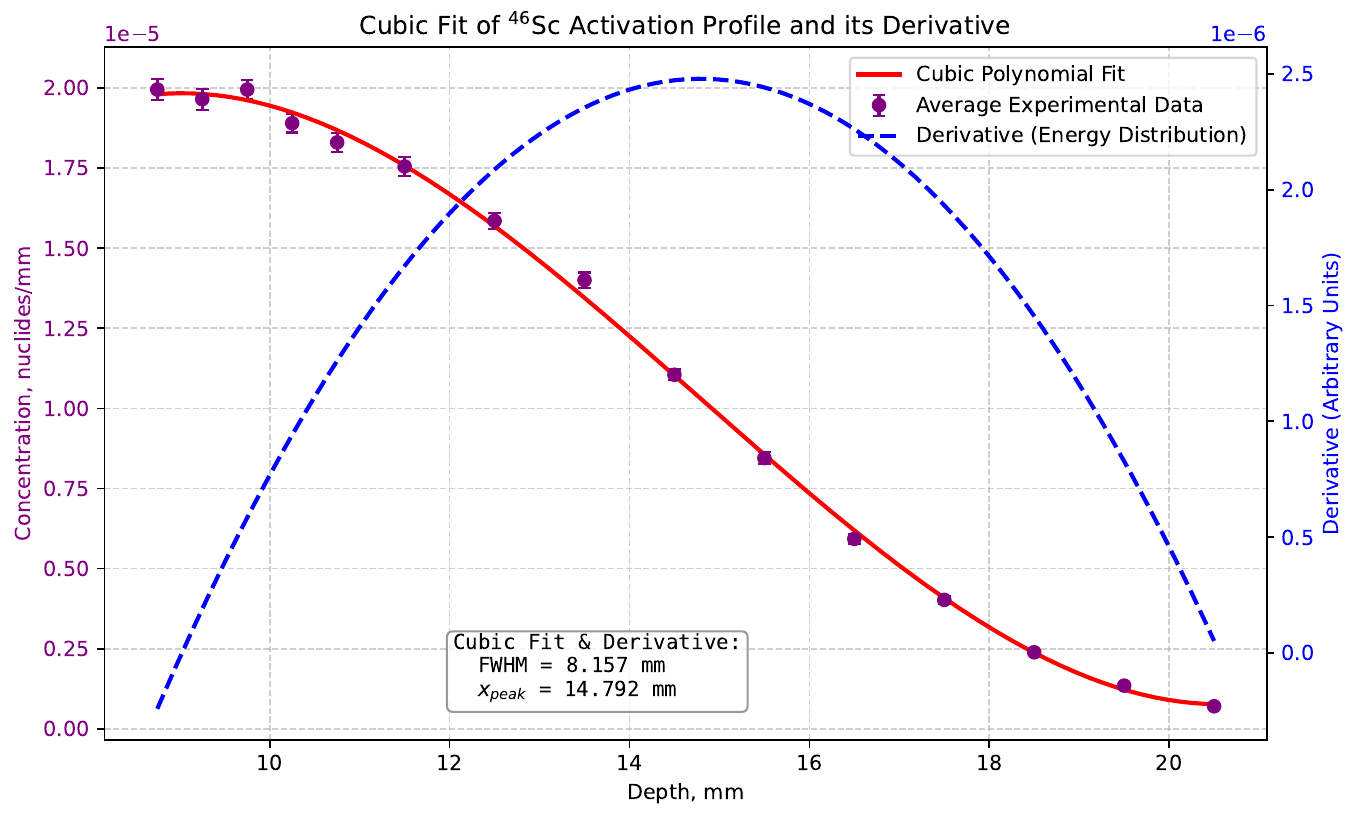}
        \caption{Cubic fit of the experimental curve.}
        \label{fig:46Sc_cube_fit}
    \end{figure}

    As already mentioned, the gaussian distribution in the JO-formalism could be replaced with any arbitrary function. For examples, a cubic fit is shown in Figure~\ref{fig:46Sc_cube_fit}.

It gives identical central $R_f=14.8$ mm, i.e. $E_c=220$ MeV/u. The FWHM = 8.157 mm corresponds to the energy spread with FWHM $\approx$ 22 MeV/u.
    
\section{Conclusion}
\label{sec:Conclusion}

 This study demonstrates the application of the JO-formalism for interpreting the activity depth profiles of projectile fragments, with $^{46}\text{Sc}$ serving as a benchmark case. Additionally, the proposed Minna-procedure provides a reliable method for determining both the production energy and the energy spread of the projectile fragments at the entrance to the target.
    
\bibliographystyle{unsrt}
\bibliography{references}

\end{document}